# Logic as a complex network

Koji Sawa

*Senior High School, Japan Women's University, 1-1-1, Nishiikuta, Tama, Kanagawa, Kawasaki, Kanagawa, 214-8565, Japan*

When we represent logical, connective implications by directed edges, the resulting set of directed edges can be regarded as a complex network. In this article, we compose a network model that represents a deductive-logic-like structure composed solely of implications. The proposed network model grows like the BA model reported by Barabási and Albert [Science **286**, 509 (1999)]. Though the BA model references the whole of the existing network when a node is added, our model references only part of the existing network. In this view, our model is more realistic than the BA model. However, it also exhibits power law characteristics.

The implicational relation "if A then C" is an ordered binary relation in formal logic. Therefore, it may be represented by a directed edge, such as "$A \to C$," instead of the frequently used representation "$A \supset C$." Thus, a real-world set composed of a huge number of implicational relations, such as "if X is a crow then X is a bird" or "if X is a kitten then X is a cat," can be regarded as a directed complex network. Though there are some formal representations of logic, as in mathematical logic [1], proof theory [2], and lattice theory [3], these representations mainly focus on algebraic structure. On the contrary, in this article, we present the network-like features of logic.

The pioneering models of complex networks are the ER model [4], WS model [5], and BA model [6]. Of these, the model proposed in this article is most similar to the BA model, proposed by Barabási and Albert. The BA model consists of two main features: growth and preferential attachment. Growth represents the step-wise evolution of a network by adding nodes and directed edges. The preferential attachment rule is employed when a new node is added to an existing network. This rule determines the probability of a new edge being formed between the new node and existing nodes according to their number of attached edges. Thus, nodes with many attached edges acquire more edges, whereas those with fewer edges struggle to acquire new edges. As a result of this preferential attachment, the BA model exhibits power law characteristics in terms of its degree distribution. The power law is sometimes referred to as being scale-free, and is a universal phenomenon displayed in the metabolic networks of organisms [7], topology of the Internet [8], directed graphs on the Web [9], and so on. The fact that the BA model demonstrates the power law implies that it is as universal as these other networks.

In general, there are two inferences in logic: deduction and induction. In this article, we consider only deductive inference, which is considered to present fewer problems than induction. One famous representation of deduction is: "All men are mortal. Socrates is a man. Therefore, Socrates is mortal." That is, deduction is broadly reasonable to most people. We formulate the process of deduction, and transform it into a network of logic based on deductive inference.

In the BA model, all information about the entire existing network is referenced when a new node is added. This is not a problem when there are few nodes or edges. However, if we have thousands of nodes or edges, this reference to the whole of the network becomes too strong and idealistic. In fact, there are some modified models that limit their reference to a local, finite part of the network [9-11]. In this article, the reference and transformation also remain local; thus, our model is more realistic than the BA model. Despite this restriction, our model also exhibits the power law in terms of degree distribution. This new result adds deductive logic (to be precise, local deductive logic that does not assume the wholeness of the network) to the collection of networks that exhibit the power law.

We now describe the construction of our model. First, we formulate deduction into a scheme. For a set $B$, we prepare a tautological binary relation $B \rightarrow B$. In logic, this binary relation is called a reflexive law; under normal conditions, there is generally no argument with this relation. Next, we bring up a subset $A$ of $B$ on the left of "$\rightarrow$," and a set $C$ whose subset is $B$ on the right of "$\rightarrow$." Thus, a new binary relation $A \rightarrow C$ is formed. This binary relation is ensured by the inclusive relation of sets or the transitive law in logic. The validity of this implicational binary relation $A \rightarrow C$ is easy to understand by substituting the sets of "cats," "kittens," and "animals" for $B$, $A$, and $C$, respectively (see Fig. 1). Roughly speaking, the transformation is a manipulation akin to "collapsing the left and expanding the right." This newly-formed binary relation is not novel, but is sound. In addition, the transformation exactly coincides with the "Socrates" example described above, where *A*, *B*, and *C* are regarded as "Socrates", "men", and "mortals," respectively. Thus, this transformation is a schematic representation of deduction.

The BA model network grows in a step-by-step manner as nodes are added. At each step, the number of edges of every node (i.e., the complete network) is checked. That is, in the BA model, it is assumed there is some viewpoint that can overlook the whole of the network, even as the network becomes huge. This assumption is not quite realistic when real-world networks are considered. Therefore, we do not adopt this assumption, and instead introduce a *local network transformation* into our model as the network grows. This local transformation proceeds as follows:

1. Choose a node *p* from an existing network.
2. Prepare a node *p'* that is a copy of *p* and a directed edge between *p* and *p'*. The direction of the directed edge is determined at random with a probability of 0.5. Regardless of its direction, the new directed edge (either $p \to p'$ or $p' \to p$) represents a tautological binary relation at this moment, because *p* and *p'* have the same representation.
3. Transform the new directed edge by the deductive transformation described above, i.e., "collapsing the left and expanding the right."
4. The set of nodes that have a directed edge to or from *p* are called the *adjacent set*. Check whether *p'* and each node in the adjacent set have an implicational relation "→" (details are described later). If *p'* and a member of the adjacent set have an implicational relation "→," form a directed edge between *p'* and that node in the natural direction.
5. Recheck the implicational relations "→" between *p* and each node of the adjacent set. If *p* retains any such relations, the directed edge between *p* and the node is maintained. Otherwise, remove the directed edge.

Steps 4 and 5 illustrate why we call the transformation "local," as only those directed edges to nodes in the adjacent set are checked. Nodes outside the adjacent set are not examined.

We implement the manipulation of "collapsing the left and expanding the right" using bit strings. A binary relation "→" between two bit strings $a_1 a_2 a_3 ... a_n$ and $c_1 c_2 c_3 ... c_n$ is defined as follows:

**Definition** $a_1\ a_2\ a_3\ ...\ a_n \rightarrow c_1\ c_2\ c_3\ ...\ c_n$ iff $a_i \leq c_i\ \forall i$, where $a_i$ and $c_i$ are either 0 or 1.

For example, the 4-bit strings 0110 → 0111 and 0101 → 1111 satisfy this definition. However, 0110 does not have a binary relation with 1101, because the third bit does not satisfy the definition. This binary relation is formed from the reflexive law of "→" by "collapsing the left and expanding the right." For example, we can consider 0101 → 1111 to be formed from 0111 → 0111 by "collapsing the left" to give 0101 → 0111 and "expanding the right" as 0111 → 1111. The former binary relation 0111 → 0111 is validated by the reflexive law of "→." This representation by bit strings also gives a representation of the deductive transformation. Figure 2 shows an example of the local transformation of a network represented by bit strings.

Using the local network transformation described above, we propose the following construction:

1. Consider $m$ nodes in the initial state. Each node is represented by an $n$-bit string. If each pair of nodes is in the state of "→," a directed edge is formed between them in the corresponding direction.
2. Choose a node at random from the existing network, and conduct the local network transformation around this node.
3. In the new node, each bit made by "collapsing the left" is set to 0 if its former value is 0; otherwise, it is set to 0 or 1 at random with a probability of 0.5. Each bit of the new node made by "expanding the right" is set to 1 if its former value is 1; otherwise, it is set to 0 or 1 at random with a probability of 0.5. This step is repeated until the left and right nodes become different.
4. If, in the process of local network transformation, we form two nodes that are the same, then they are linked to each other by two reciprocal directed edges.
5. Steps 2–4 are iterated until we reach a predefined number of nodes $N$.

We conducted a trial with $m = 2$, $N = 10000$, and 14-bit strings. The initial nodes are $p_1$ = 00000001111111 and $p_2$ = 00000001111111; thus, there are two reciprocal directed edges $p_1 \rightarrow p_2$ and $p_2 \rightarrow p_1$. As shown in Fig. 3, both the out-degree and in-degree exhibit power-law distributions of the node degree. Figure 4 shows the evolution of the clustering coefficient and average path length with respect to $N$. The clustering coefficient remains constantly high, and the average path length is proportional to $\log(N)$. Note that the edge directions are ignored in calculating the clustering coefficient and average path length; that is, the networks are regarded as undirected networks.

These results show that, like many other networks, deductive logic has universal structure. As logic is used by every healthy subject, this would appear to be valid. That is, logic is one of the most abstract computational models to have been formulated from the real world. If logic is not universal, it would never have become so popular.

The network in the model is updated step by step, including existing nodes and directed edges. At first, the change in the representation of a node (the change in its bit string) may be unexplainable, as may the addition or elimination of a directed edge. However, these aspects of the model correspond to a reconsideration of our recognition of the real world. We sometimes change our view of the world drastically—the change of representation of a node and the addition/elimination of a directed edge represent this drastic change. Logic that has already been formalized is static and does not vary. However, our logical judgment in the real world is dynamic and pliable. The model in this article is a representation of the latter. Such a reconsideration has been argued in the area of artificial general intelligence [12]. The most important aspect of our model is the sustainment of the power law, no matter how the network is updated.

**Acknowledgement** This work was carried out under the supports by the Cooperative Research Project Program H25/A12 of the Research Institute of Electrical Communication, Tohoku University.

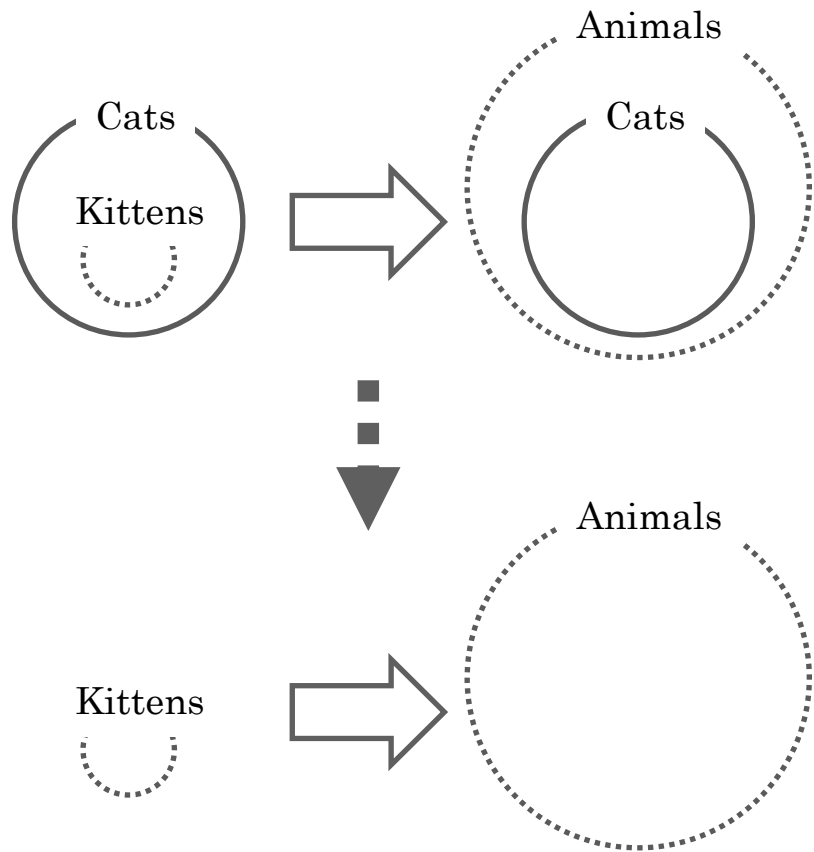

Figure 1: Forming implicational relations. This transformation is represented as "collapsing the left and expanding the right" from the tautological binary relation "cats → cats."

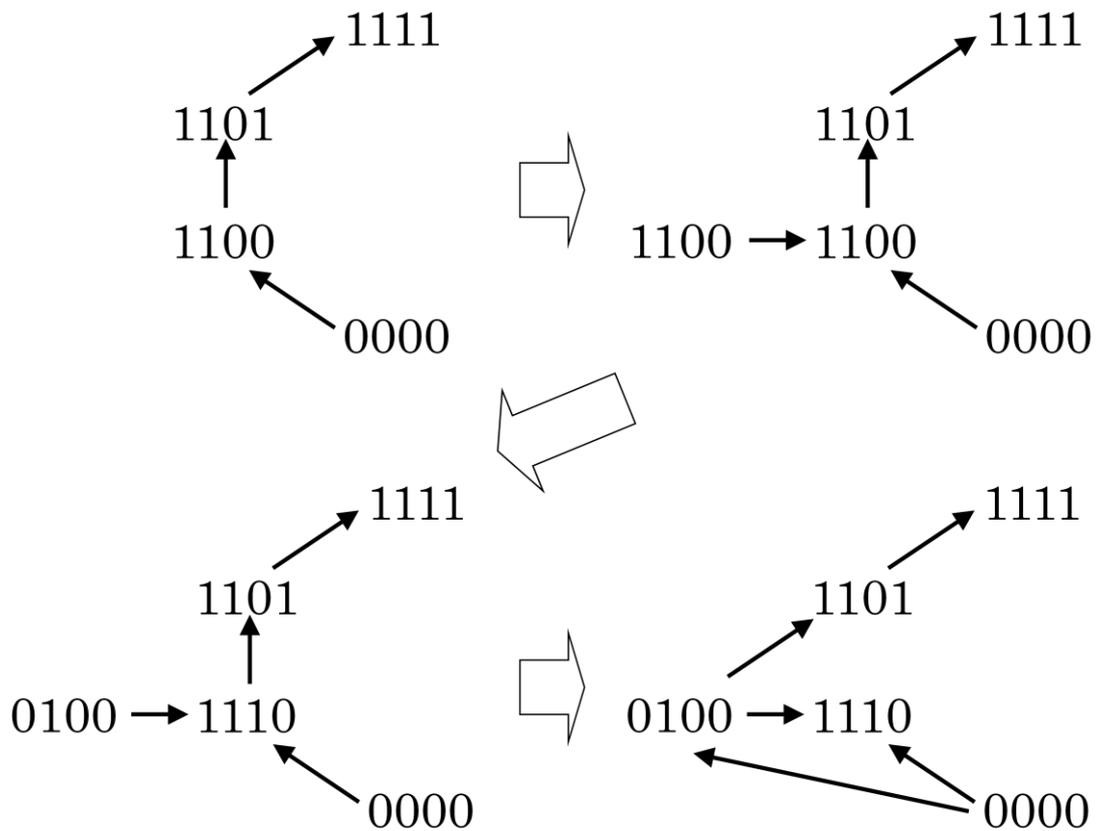

Figure 2: An example of local network transformation represented by bit strings. From the initial network (upper left), a node *p* = 1100 is randomly chosen and copied as *p*'. Thus, a tautological binary relation is formed (upper right). The tautological binary relation is transformed by the deductive transformation of "collapsing the left and expanding the right" (lower left). The relations between either *p* = 0100 or *p*' = 1110, and the existing nodes 0000 and 1101 that have the relation "→" with *p* are checked. If two nodes are related by "→," a directed edge is added/retained; otherwise, the edge is removed/not added (lower right). By this local network transformation, the existing relation 1110 → 1101 is removed. The node 1111 is not completely referenced by this local network transformation. This explains the meaning of the "local" network transformation.

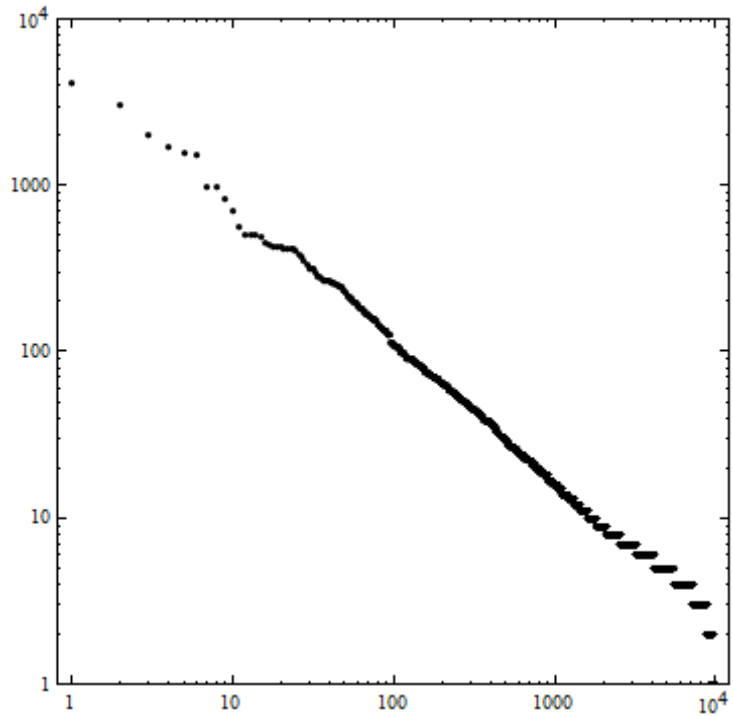

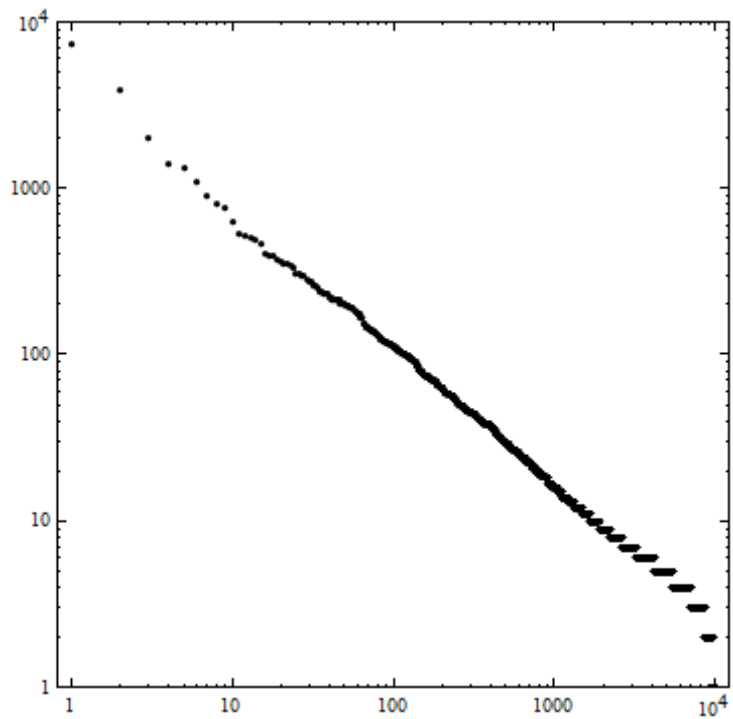

Figure 3: Distributions of node degrees. (a) Out-degree and (b) in-degree. Each graph shows the relation between the degree (vertical axis) and the rank ordered by the degree (horizontal axis). Both graphs are double logarithmic plots.

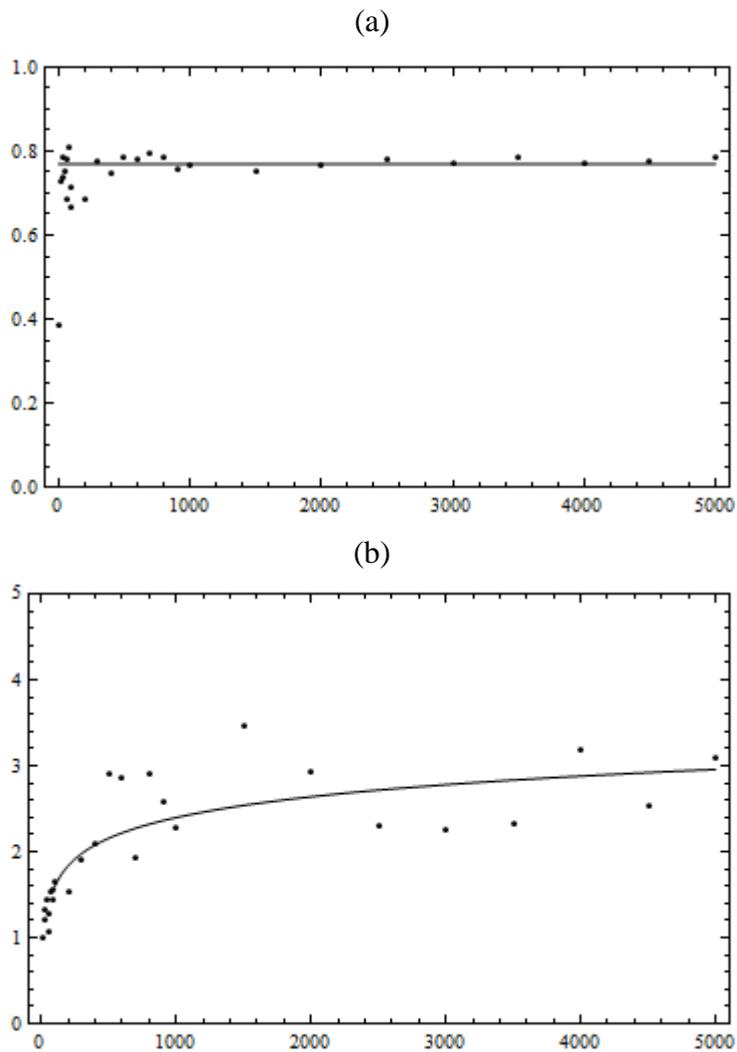

Figure 4: (a) Evolution of clustering coefficient with respect to *N* for values of *N* = 10, 20, ..., 100, 200, ..., 1000, 1500, 2000, ..., and 5000. The straight line is positioned at 0.77. (b) Evolution of average path length with respect to *N* for the same values of *N*. The curved line is the plot of $(4 \log_{10}N)/5$.